# A Secured Health Care Application Architecture for Cyber-Physical Systems

**Jin Wang[1,2], Hassan Abid[2], Sungyoung Lee[2], Lei Shu[3] and Feng Xia[4]**

[1] *School of Computer and Software, Nanjing University of Information Science & Technology, 210044, Nanjing, China (e-mail: wangjin@oslab.khu.ac.kr)*
[2] *Department of Computer Engineering, Kyung Hee University, 449-701, Yongin City, South Korea (e-mail: {hassan, syle}@oslab.khu.ac.kr)*
[3] *Department of Multimedia Engineering, Osaka University, 577-8505, Osaka, Japan (e-mail:Lei.shu@ieee.org)*
[4] *School of Software, Dalian University of Technology, 116620, Dalian, China (e-mail: f.xia@ieee.org)*

**Abstract:** Cyber-physical systems (CPS) can be viewed as a new generation of systems with integrated control, communication and computational capabilities. Like the internet transformed how humans interact with one another, cyber-physical systems will transform how people interact with the physical world. Currently, the study of CPS is still in its infancy and there exist many research issues and challenges ranging from electricity power, health care, transportation and smart building etc. In this paper, an introduction of CPeSC3 (cyber physical enhanced secured wireless sensor networks (WSNs) integrated cloud computing for u-life care) architecture and its application to the health care monitoring and decision support systems is given. The proposed CPeSC3 architecture is composed of three main components, namely 1) communication core, 2) computation core, and 3) resource scheduling and management core. Detailed analysis and explanation are given for relevant models such as cloud computing, real time scheduling and security models. Finally, a medical health care application scenario is presented based on our practical test-bed which has been built for 3 years.

***Keywords:*** cyber-physical systems, cloud computing, health care, wireless sensor networks, security.

## 1. INTRODUCTION

Cyber-physical systems (CPS) have drawn the attention of many researchers from both academic and industrial field recently and they are viewed as the next computing revolution. A CPS can be viewed as a new generation of system with integrated communication and computational capabilities which can interact with the physical world. Just as the internet transformed how humans interact with one another, cyber-physical systems will transform how people interact with the physical world [Lee 2006; 2008].

The U.S. National Science Foundation (NSF) has identified CPS as a key research area in 2008 and it has also been listed as the No. 1 research priority by the US President's council of advisors on science and technology. CPS has wide potential applications in defence, transportation, industrial automation, energy, health care and agriculture etc. Taking BMW as an example, it can be viewed as a network with hundreds of sensors and actuators which interact with their surrounding physical environment and provide decision support to some cyber systems via communication, computation, control and resource management etc. Here it can be seen that CPS can transform our world from a physical system to a cyber system with faster response, more accuracy as well as higher QoL (quality of life) etc.

There are many research issues and challenges in terms of designing such CPS since it has a very broad spectrum of systems types and the research is multi-disciplinary [Lee 2008]. Some of the existing problems cover areas ranging from electric power grid, environmental protection to communication and monitoring systems. To list a few of them, the understanding of CPS (like the boundary, abstraction etc.), the architecture and platform design, the hardware and software design with QoS requirement, the application and interface with human etc. all belong to such existing problems category.

Here the details of those problems and challenges will not be given and two examples are provided. For example, CPS should meet real time requirement with security and reliability. Since physical environment tends to be time varying with dynamic network topology, minor changes should be reported and handled in a timely manner. Another example is the requirement of scalability and interoperability between different systems which is also a nontrivial task.

CPS can be used to many medical health care applications with more powerful communication, computation and security capabilities than today's counterparts. For example, various types of sensors can collect the information from patients at home and then communicate with a third party cloud server which has more powerful computation capability. In the mean time, the doctors in hospital can remotely monitor the patient's physical condition and give suggestions or prescriptions. The main focus in this paper is to build a secured health care application architecture for CPS by integrating sensing, communication, computation and security core together. Research issues such as cloud computing, real time scheduling and resource management and security etc. are analyzed.

The contribution in this paper includes:

1) A comprehensive survey about CPS related research work in communication and network area is provided, as can be seen from Fig. 1 and the references;

2) The CPeSC[3] (cyber physical enhanced secured WSN-integrated cloud computing for health care) architecture is proposed which is consisted of three components, namely 1) communication and sensing core, 2) computation and security core, and 3) real-time scheduling and resource management core;

3) A medical health care application scenario is presented based on our practical test-bed which has been built for 3 years.

The remainder of the paper is organized as follows. Section 2 presents some related work about CPS. Section 3 gives an overview of the CPeSC[3] architecture. Section 4 presents detail explanation of various components and models inside CPeSC[3] architecture such as cloud computing model, real time scheduling model and security model etc. In Section 5, a medical health care application scenario is presented to deepen the understanding of the CPeSC[3] architecture. Section 6 concludes this paper and gives some future work.

## 2. RELATED WORK

### *2.1 CPS Related on-going Research Groups*

The study of CPS is still in its infancy. The 1st international CPS workshop (WCPS) was sponsored by IEEE and the U.S. National Science Foundation in Beijing China 2008, and the 2nd one was held in Quebec Canada 2009. It is found that the papers therein are mainly dealing with CPS secure issues as well as modelling problems [Jing et al. 2009]. Recently, some papers about middleware [Zhang et al. 2008], congestion control [Ahmadi et al. 2010], QoS (quality of service) management [Xia et al. 2008] as well as sensor related applications [Xia et al. 2010] also appear.

Before further explanation about detailed related work, some of the CPS related on-going work in communication and network area at some research labs in the world is presented, as is shown in Fig. 1.

From Fig. 1 some common research issues can be found such as 1) wireless networks as basic infrastructure (like sensor network, body area network etc.); 2) security as a critical and challenging issue; 3) applications (like health care, middleware, control issues and data mining etc.). The interested authors can refer to Table 1 for more information.

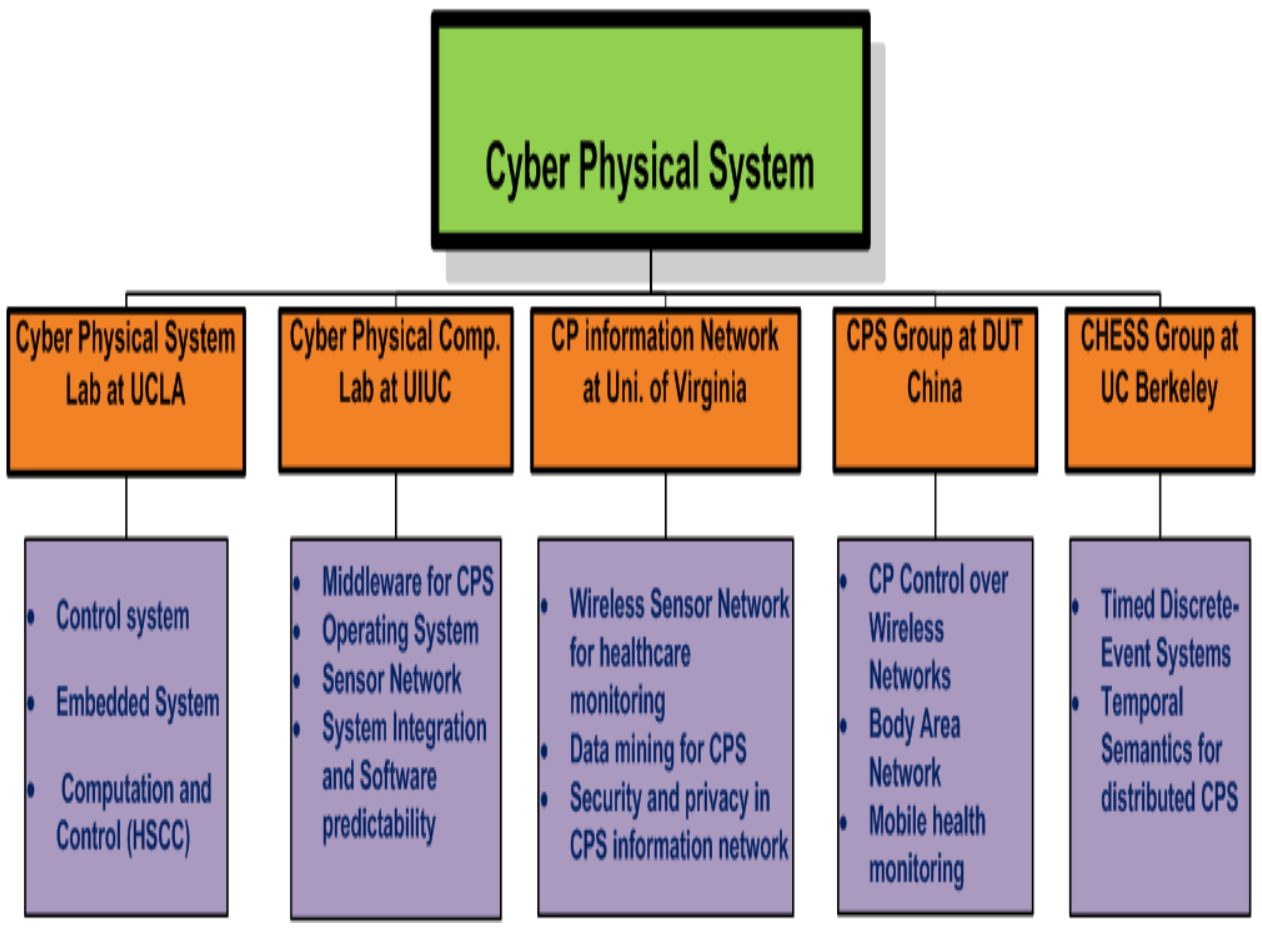

Fig. 1. CPS related on-going work at some research labs.

**Table 1. CPS related research lab link address**

| Lab | Link address |
|---|---|
| UCLA | http://www.cyphylab.ee.ucla.edu/Home/cyber-physical |
| UIUC | http://www.cs.uiuc.edu/homes/zaher/cyberphysical/index.html [Pham et al. 2010] |
| Virginia Univ. | http://www.cs.virginia.edu/~stankovic/cps.html [Rajkumar et al. 2010] |
| DUT, China | http://cpschina.org and http://fengxia.net/ [Xia et al. 2008, 2010] |
| UC Berkeley | http://chess.eecs.berkeley.edu/ [Lee 2006; 2008; Eidson et al. 2010] |
| MIT | http://ares.lids.mit.edu/index.php?option=com_frontpage&Itemid=1 |
| Penn State Univ. | http://mlab.seas.upenn.edu/ |
| Missouri S&T | http://web.mst.edu/~sendecomp/projects.html |

### *2.2 CPS Related Research Work*

Next, some CPS related work is presented in a similar way as Fig. 1 which covers survey papers, application issues, security problems and finally some sensor related applications.

#### *2.2.1 Survey Work*

Among the survey work, the author in [Lee 2008] listed some requirements and challenges in designing CPS and the conclusion was that to realize the full potential of CPS, to rebuild computing and networking abstractions was necessary. The authors in [Sha et al. 2008] present some challenges and promises of CPS such as real-time system abstractions, robustness and security, system QoS composition as well as trust etc. Also, they showed a medical device network example to provide better understanding of challenges for CPS. Three key challenges for securing cyber-physical systems are discussed in [Cardenas et al. 2009] and the authors in [Rajkumar et al. 2010] pointed out some challenges and vision of CPS from a broad view ranging from power grid, scientific discovery, and emergency evacuation to assistive devices. The author in [Lee 2006] proposed some pioneering work for CPS by examining the foundations of computing, identifying some promising research technologies and finally listing some research directions. The author in [Gupta et al. 2011] first classified CPS resource management algorithms into four types and then discussed each of their pros and cons with respect to data centres and body sensor networks.

#### *2.2.2 CPS Applications*

Among CPS applications, the authors in [Zhang et al. 2008] proposed a configurable component middleware services for admission control and load balancing in distributed cyber-physical systems. The authors in [Jing et al. 2009] developed a qualitative and quantitative understanding of the dependability in CPS which was characterized in terms of attributes. Simulation results also validated the relevant models. A novel spatial aggregation congestion control mechanism for accurate estimation of spatio-temporal phenomena was proposed for CPS in [Ahmadi et al. 2010]. They adopted different granularities of aggregation in transporting spatio-temporal data from nodes to a base station and the aggregation granularity was chosen locally based on the contribution of the transmitted data at the receiver. The protocol was implemented on Micaz motes and better performance was evaluated through test bed experiments. The authors in [La et al. 2010] applied service-oriented architecture to the CPS which was called service-based CPS. The three-tier architecture was proposed with detailed design methods dealing with design challenges of CPS.

The authors in [Xia et al. 2008] examined the main characteristics and challenges of wireless sensor/actuator networks which play an essential role in CPS. As one of the challenges, network quality of service management in the context of CPS was studied and a feedback scheduling framework was proposed with an illustrative example to shown the effectiveness of their method.

Based on their previous work in [Lee 2006; 2008], the authors in [Eidson et al. 2010] used an extension of Ptolemy II framework as a coordination language for the design of distributed real-time embedded systems. By combining PTIDES (programming temporally integrated distributed embedded systems) with modal models, they illustrated timed mode transitions which enable time-based detection of missing signals to drive mode changes in the operation of common industrial applications.

#### *2.2.3 Privacy and Trustworthiness Applications*

Among privacy and trustworthiness applications, the authors in [Pham et al. 2010] derived fundamental bounds on privacy achievable in future human-centric CPS as well as an optimal trade-off between privacy and community reconstruction accuracy. In [Tang et al. 2010], the authors proposed a Tru-Alarm method which can find out trustworthy alarms and increase the feasibility of CPS. Once the locations of objects causing alarms were estimated, an object-alarm graph would be constructed and trustworthiness inferences would be carried out based on link information in the graph. Simulation results validated that their method can effectively filter out noises and false information and guarantee the accuracy of missing any meaningful alarms.

#### *2.2.4 Sensing related Applications*

Among sensing related applications, the authors in [Xia et al. 2010] studied some of the major QoS challenges raised by wireless sensor and actuator networks like resource constraints, platform heterogeneity, dynamic network topology and mixed traffic etc. They analyzed the behaviour of wireless channels via simulations based on a realistic link layer model and proposed a solution which took advantage of existing prediction algorithms. Simulation results were provided to evaluate the performance of several prediction algorithms.

#### *2.2.5 Our Distinguish Feature*

It is worth noting that in this CPS related research work, other research fields are not provided such as electricity energy system, intelligent building, transportation etc. The focus of this paper mainly includes communication, networking and sensing related research area for CPS.

We also study two other architectures which are similar to ours. In UIUC research lab, the researchers mainly addresses challenges in integration of sensor, personal computing and networking devices with autonomic computing platforms, clusters, grids and web. The architecture in [Rajkumar et al. 2010] consists of four information networks including medical, hospital, living and monitoring network. They investigate in the area of cyber-physical information network with an emphasis on event detection, reliable data analysis, spatiotemporal analysis and privacy and security of cyber-physical information network. Their major contributions are related to evaluating height for biometric identification, sleep monitoring and few other body sensor network applications.

In contrast to their work above, our work focus on leveraging cloud computing technology and its integration with smart health care application to improve the performance of existing $SC^3$ system while taking into account cyber-physical concepts. In this paper a $CPeSC^3$ architecture is proposed with various components and modules which is a summary of our last 3 years' research work. The design of each components and models are explained in detail such as cloud computing and security modules, the communication, computation, scheduling and resource management cores inside $CPeSC^3$. A health care application scenario is also demonstrated to deepen the understanding of our proposed $CPeSC^3$ architecture for CPS.

## 3. OVERVIEW of $CPeSC^3$ ARCHITECTURE

### *3.1 $SC^3$ System*

The CPeSC³ architecture is an improved version of our previous Secured WSN-Integrated Cloud Computing for U-life Care (SC³) system which has been developed for 2 years.

Fig. 2 shows an overall concept of SC³ system which is consisted of three main modules, namely ubiquitous sensors & wireless sensor networks (WSNs), cloud computing and users modules. It can be viewed as a coarse prototype of CPS since it comprises:

The communication core between WSNs, cloud computing and user modules;

The computation core inside WSNs and cloud computing modules, e.g., activity recognition, inference engine;

The resource scheduling and management core among users and different databases (DB).

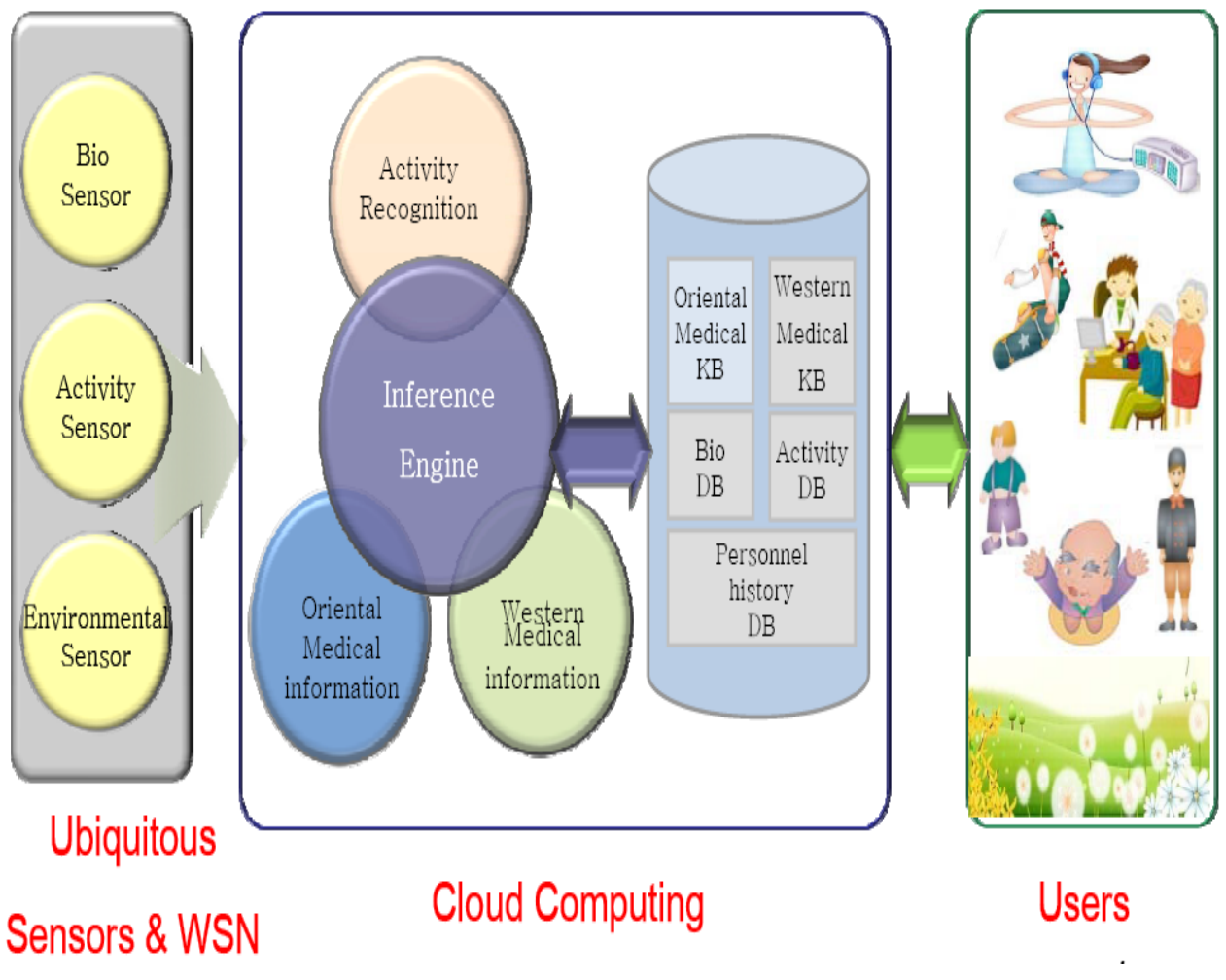

Fig. 2. Overall concept of SC³ system.

Due to the space limitation of this paper, further detail explanations are not given here. People can refer to the paper [Hung et al. 2010] to see more information.

### *3.2 CPeSC³ System*

Based on Fig. 2, a Cyber-Physical enhanced SC³ (CPeSC³) architecture is proposed in Fig. 3 by:

1) Introducing the concept of communication core and computation core from CPS;

2) Introducing a security core module;

3) Introducing some applications for users at upper layer.

It can be seen in Fig. 3 that the CPeSC³ architecture can be viewed as a representative CPS which comprises sensing, communication, computation, security and cloud computing cores. In the next section the main components and models under a modularized logical CPeSC³ architecture is presented.

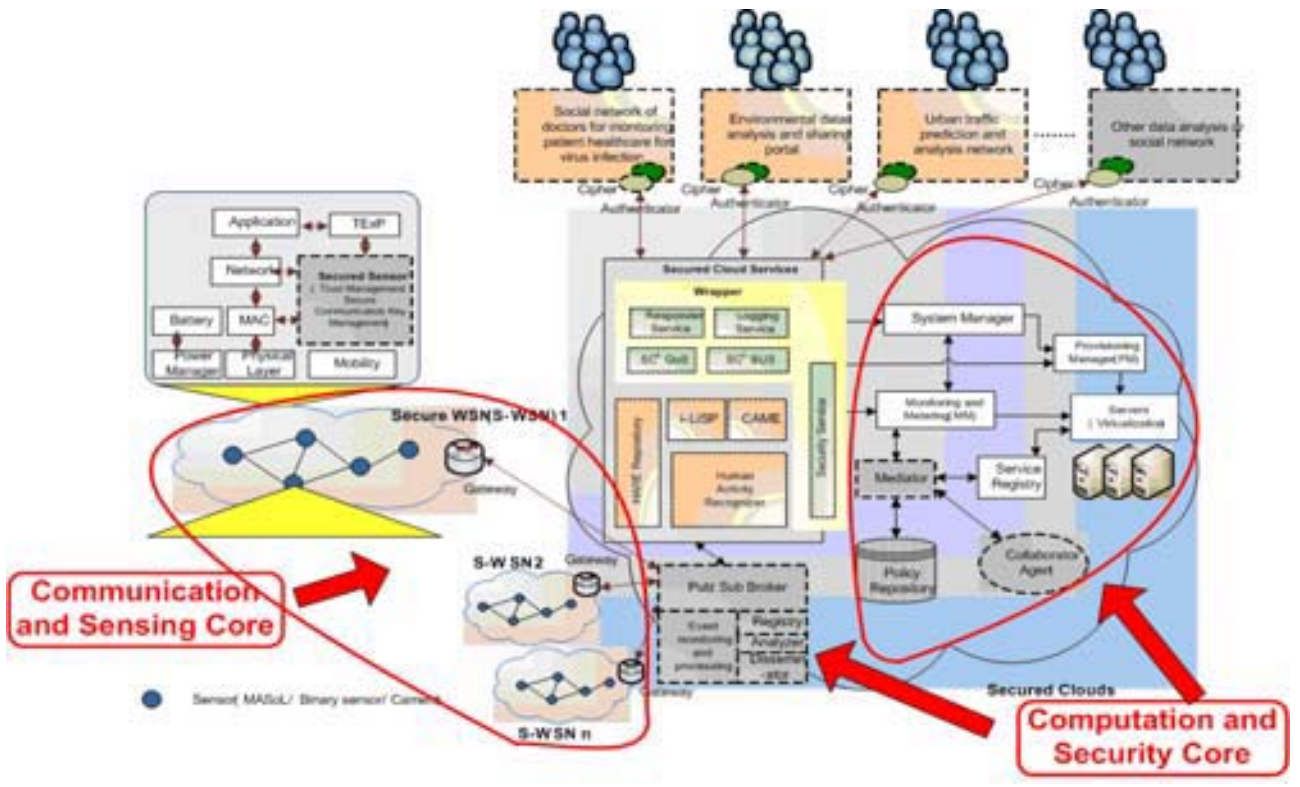

Fig. 3. Overall architecture of CPeSC³.

## 4. DETAIL ANALYSIS of CPeSC³ ARCHITECTURE

Fig. 4 gives a modularized logical architecture of CPeSC3 which divides the functionalities of communication, sensing, computation cores etc. into smaller modules.

The sensed data is either human health data or data to be used for detection of human activities for health care services. The sensors are either attached to a person or to the walls in the home environment. The video-based approach is based on images collected from camera, extracting the background to get the moving object and inferring activities such as walking, sitting, standing, falling down, bending, jacking, jumping, running etc. Image-based authentication and activity-based access control mechanism can be adopted to enhance security and flexibility of user's access.

It is worth noting that the security core is not a separate module but can be applied to both communication and computation modules. Also it can be seen that the cloud computing service is incorporated into the computation core for economical reason. In the next several sub-sections, each of the components or modules is explained in detail.

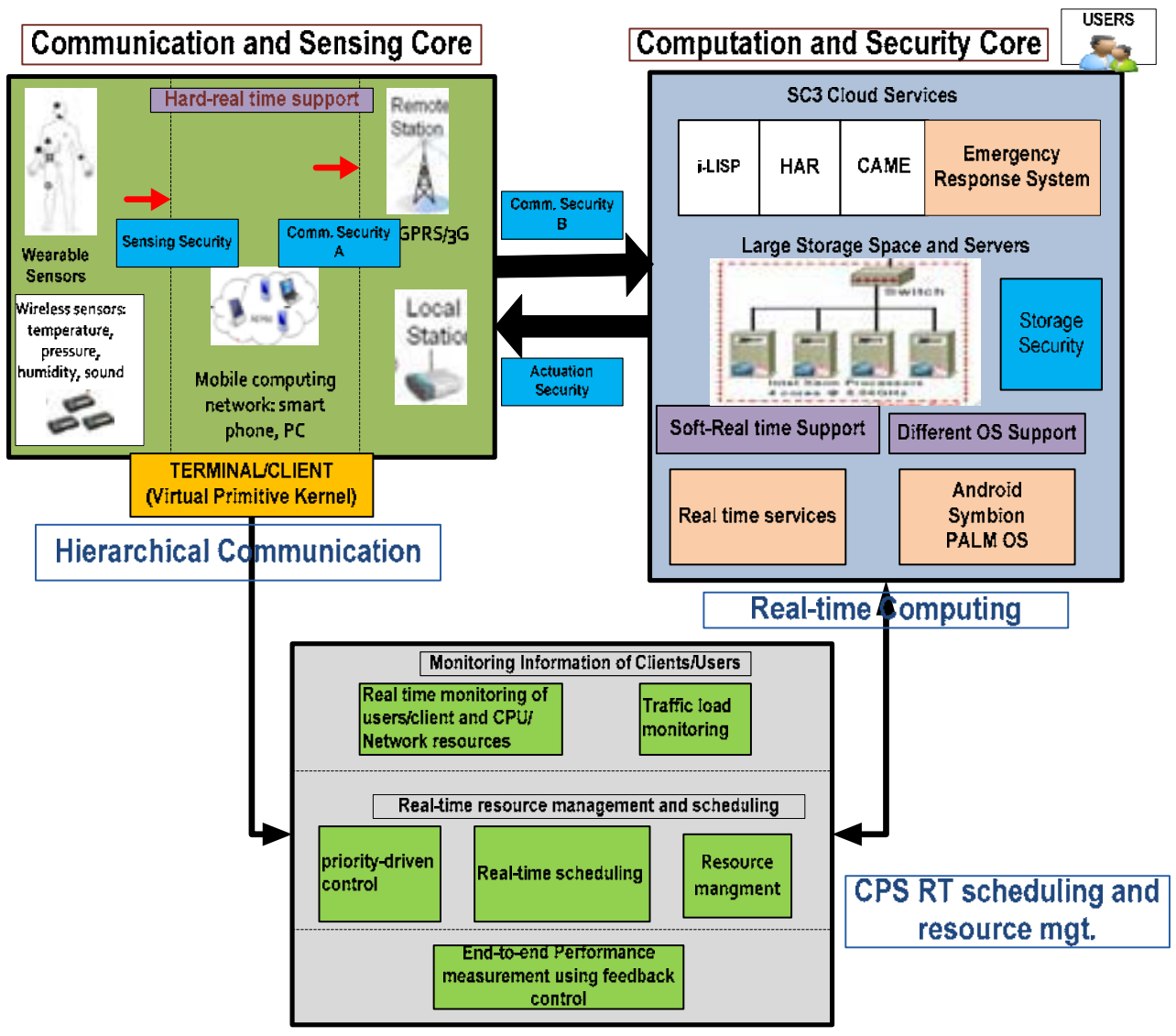

Fig. 4. Modularized logical CPeSC3 architecture.

*4.1 Communication and Sensing Core*

The current technologies that integrate computations and interactions with physical world typically emanate from the fields of embedded systems and real time systems. Wireless sensor networks are different from their traditional ad hoc wireless counterparts in that they have a large scale, higher density and smaller devices, and tighter interactions with the physical environment. So the key requirement is to sustain for long lifetime on limited power supplies for WSN. Meanwhile, due to the criticality of CPS applications, many computation and communication tasks must be finished within timing constraints to avoid undesirable or even catastrophic consequences. Thus to ensure real time support in the large-scale wireless sensor network is also an important and challenging research issue.

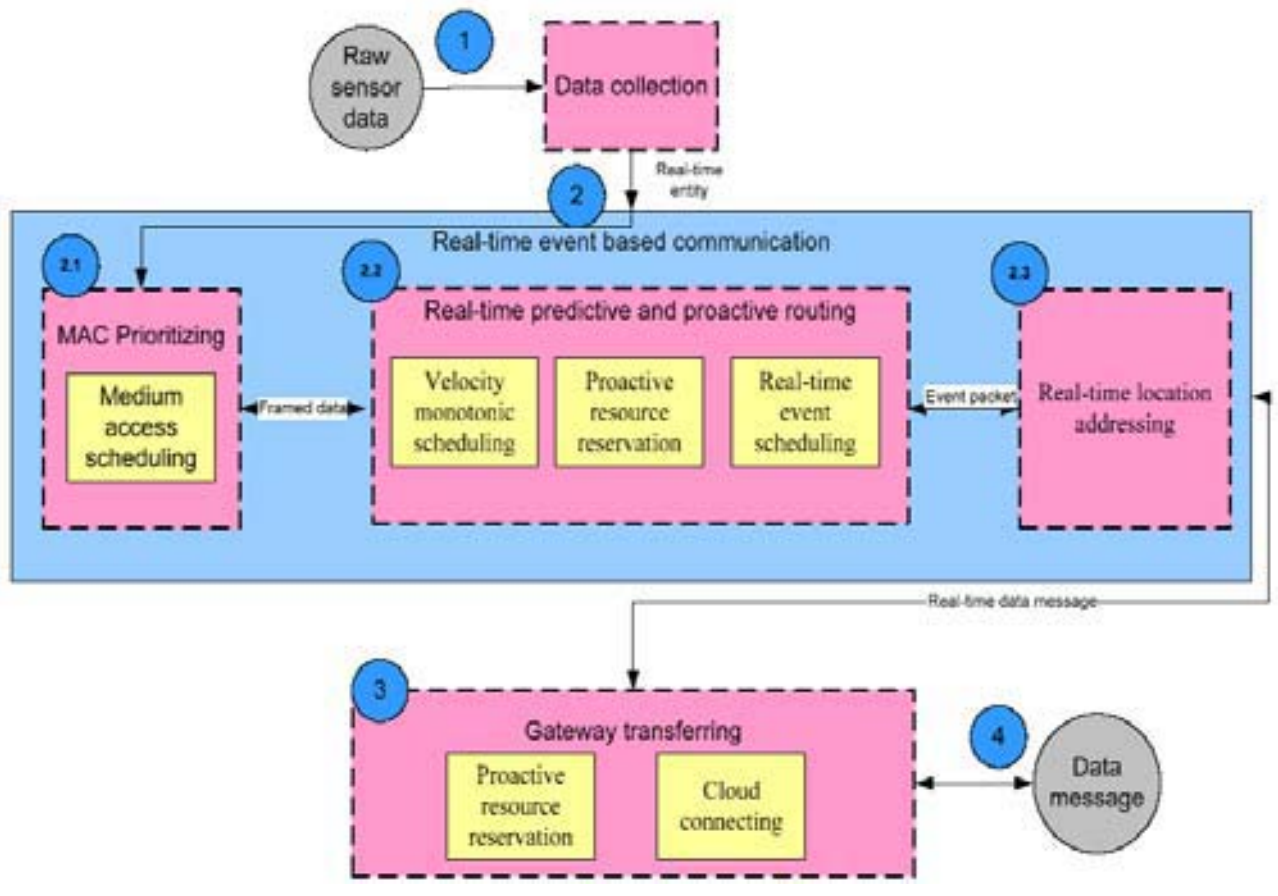

Fig. 5. Data flow chart in communication and sensing core.

Fig. 5 shows the data flow chart from raw sensor data to data transferred by gateway and stored in certain database for further processing. Some critical research issues include medium access scheduling, routing, real time location and/or addressing etc.

The hierarchical communication (see Fig. 4) can easily divide the network into different tasks with reconfigurable mapping and pipeline techniques. In this way, the communication core can efficiently enhance continuous and timely monitoring of patient for CPS.

Communication in a sensor network can be hard real-time or soft real-time communication system (see Fig. 4). The real time communication module is to build a real time abstraction layer that needs novel distributed real time computing and real time group communication methods in wireless CPS with mobile components under dynamic network topologies.

*4.2 Computation Core*

From Fig. 3 above, it can be seen that the secured cloud services can provide upper layer users with applications such as social network of doctors for monitoring patient healthcare, environmental data analysis, urban traffic prediction and analysis network etc.

In fact, cloud computing has pivotal role in providing high performance computing, integrating mobile devices with cloud and supporting different types of operating system (OS) platforms etc. Fig. 6 shows the cloud computing model in computation core as a key module where three sub-modules of SaaS (Software-as-a-Service), PaaS (Platform-as-a-Service) and IaaS (Infrastructure-as-a-Service) provide basic services for it.

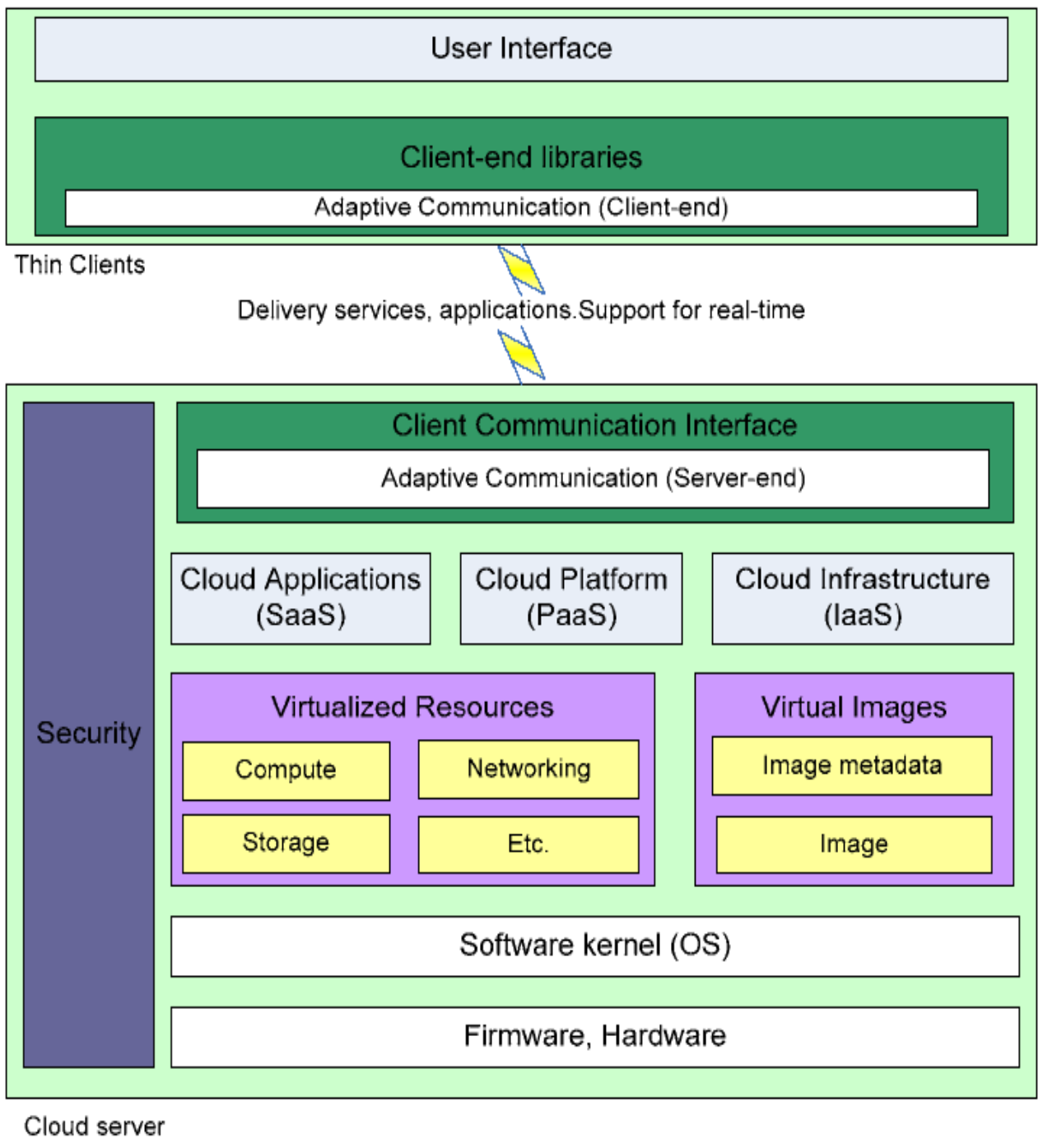

Fig. 6. Cloud computing model in computation core.

From Fig. 6 it can be seen that the cloud computing model comprises of two sub-components named cloud server and thin clients. In the meantime, it can support real time delivery services and applications. Interested readers can refer to our lab's website for more information.

*4.3 Real-time Scheduling and Resource Management*

The traditional server centric approach has been transformed into network or cloud centric approaches. With the advent of such new approaches, data centres have been transformed into server virtualized networks that are supported by hardware assisted virtualization. To meet the ever increasing demand from users, there is increasing need of new algorithms for real time scheduling and resource management which are configurable from economic point of view. In Fig. 7, a three layered reference model for real time scheduling and monitoring is presented. It interacts with communication and computation cores, as can be seen in Fig. 4.

On the bottom layer, end-to-end performance measurement can be done inside client/server or client cloud based on the real-time data collected by different sensors and cameras. Some of the functions include managing virtual resources, creating VMs (virtual machines) with software necessary, scheduling jobs to VMs, improving computation time, reducing transmission time and meeting the deadline and QoS requirement.

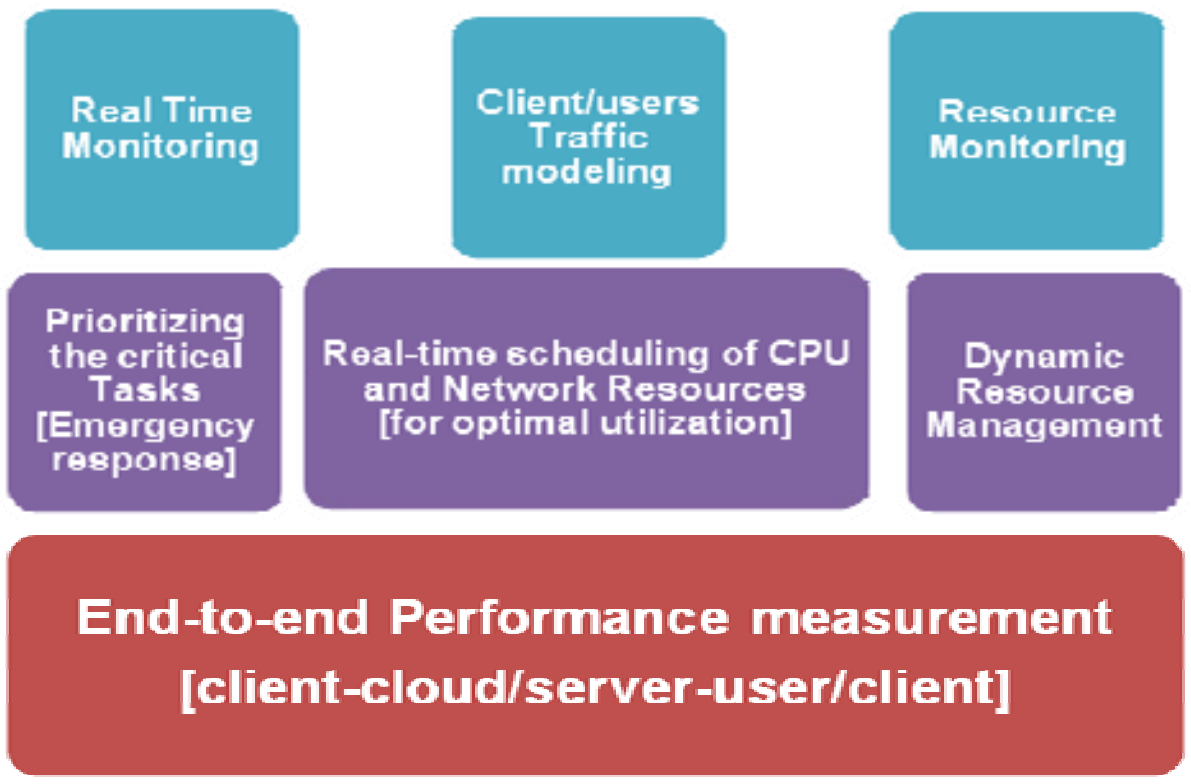

Fig. 7. Reference model for real time scheduling and monitoring.

On the middle layer, prioritizing techniques is adopted in order to deal with emergent responses like medical health care, transportation accident, electricity blackout or brownout etc.

It is worth noting that Fig. 7 covers real time scheduling (middle layer) and monitoring (top layer) of resources such as communication resources (like network bandwidth, energy consumption), computational resources (like CPU, memory utilization) and other resources from server and client side.

*4.4 Security Core*

In Fig. 8 it can be seen that the CPeSC$^3$ security model comprises two parts: sensing & communication security as well as storage & actuation security.

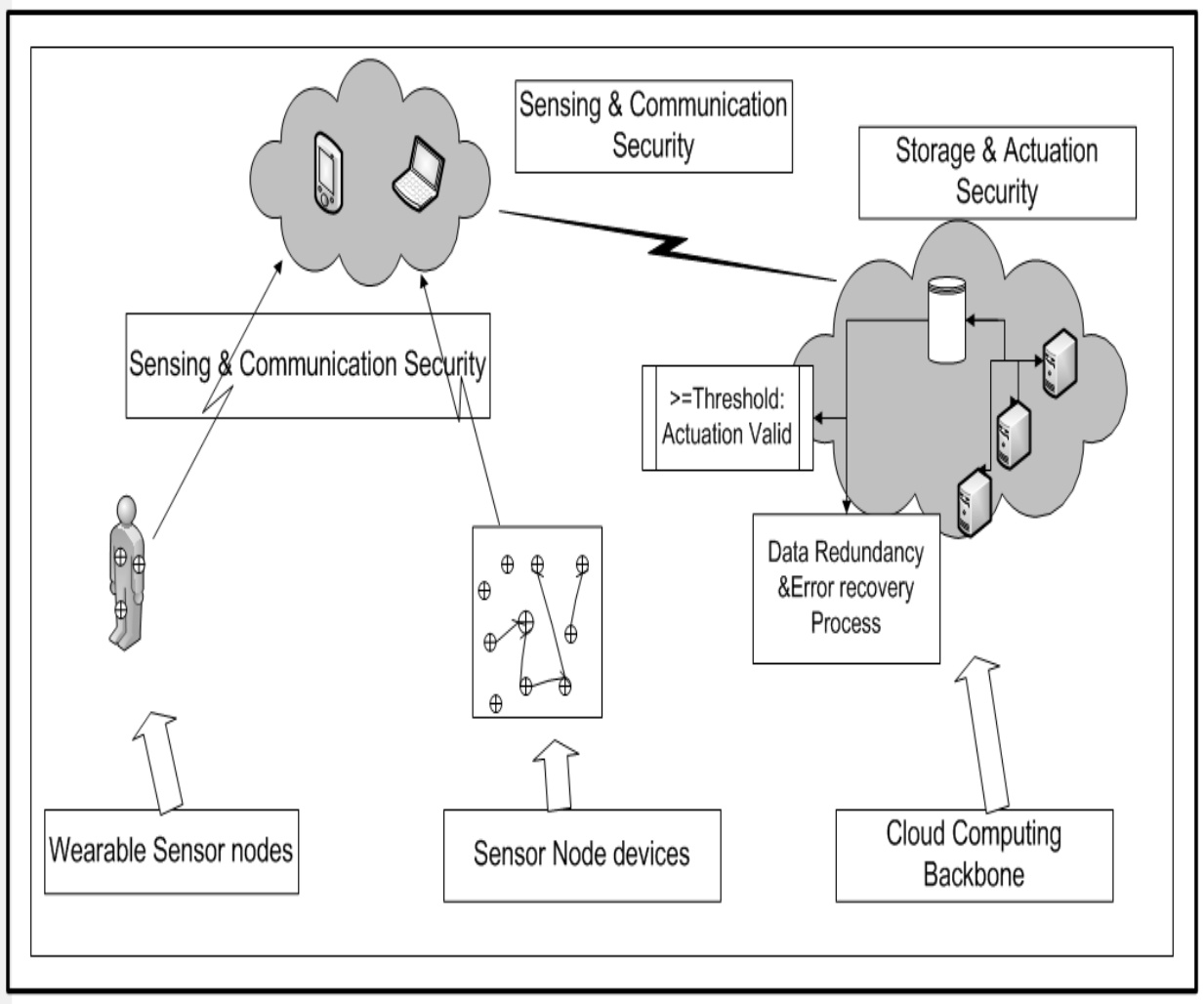

Fig. 8. CPeSC$^3$ security model.

To deal with sensing & communication security, a distributed checkup mechanism on mobile computing side is deployed as an effective defence after encrypting the actual sensed data with a shared or symmetric key. If any misbehaviour report is obtained on the checkup system, a drop will be called which results in protecting the other party of the system from being affected or attacked.

On identity privacy, concatenation is applied between source sensor node and random number to provide protection against attacks and then encrypt them to provide anonymity of the source node against some attacks.

It is assumed that the gateway or base station is a central command authority which cannot be compromised by an attacker. In order to provide protection against en-route attacks from traffic analysis or fabrication during transfer from one node to another, a secure communication model can be built which can help to establish hybrid key (asymmetric key and symmetric keyed hash function) scheme. On this basis, asymmetric key (between the sensor node and the base station) is used only for hiding the sensor node identity (anonymity) while the symmetric keyed hash function is used to protect the whole actual sensed data. The scheme is also resilient to the second type of node compromised, where an attacker injects the nodes in the network with the false identities. In this case, the base station will be able to detect this attack through a failure verification of our anonymity mechanism.

It is not easy for an adversary trying to compromise a sensor node due to the infeasible computational properties of keyed hash function. This makes it extremely difficult for an adversary to retrieve the necessary keys to decrypt or gain access to the original message. This also provides a simple resistance in the case of nodes compromising as the key established between non compromised nodes remains confidential.

To deal with storage & actuation security, a file retrieval and error recovery based mechanism is deployed to detect any unauthorized data modification and corruption due to server compromise or random failures.

Once data has been processed, it may be required to be stored over time for future access on cloud storage. Any misbehaviour or altered of this stored data can lead to errors within the whole system. Data redundancy can be employed by using a technique of erasure correcting code to further tolerate faults or server crash on the cloud point of view as the data grows in size.

To valid an actuation process, a set of rules can also be applied to ensure that no actuation can take place without an appropriate authorization during an active mode of operation.

## 5. A MEDICAL HEALTH CARE APPLICATION SCENARIO

Fig. 9 gives our medical health care application scenario which comprises of three main sites, namely home, hospital and office environment. The purpose of this developed application is to improve the QoL of elderly person (or patients) whose children are working in the office at daytime and to save medical cost.

In normal situation, the daily physical and/or medical information of an elderly person is collected (by camera, biosensor, micaz etc.) and stored in a third party cloud (like the main lab) via an in-home WSN-Cloud Gateway. The child in the office and hospital doctors (or nurses) can regularly check such medical record and give some suggestions and/or prescriptions via wired or wireless connection to the cloud in an authenticated way. In urgent situation (like an elderly person falls down), such emergent information will be sent in a timely manner to both the doctors and their family members so that immediate actions can be taken to help the fallen people.

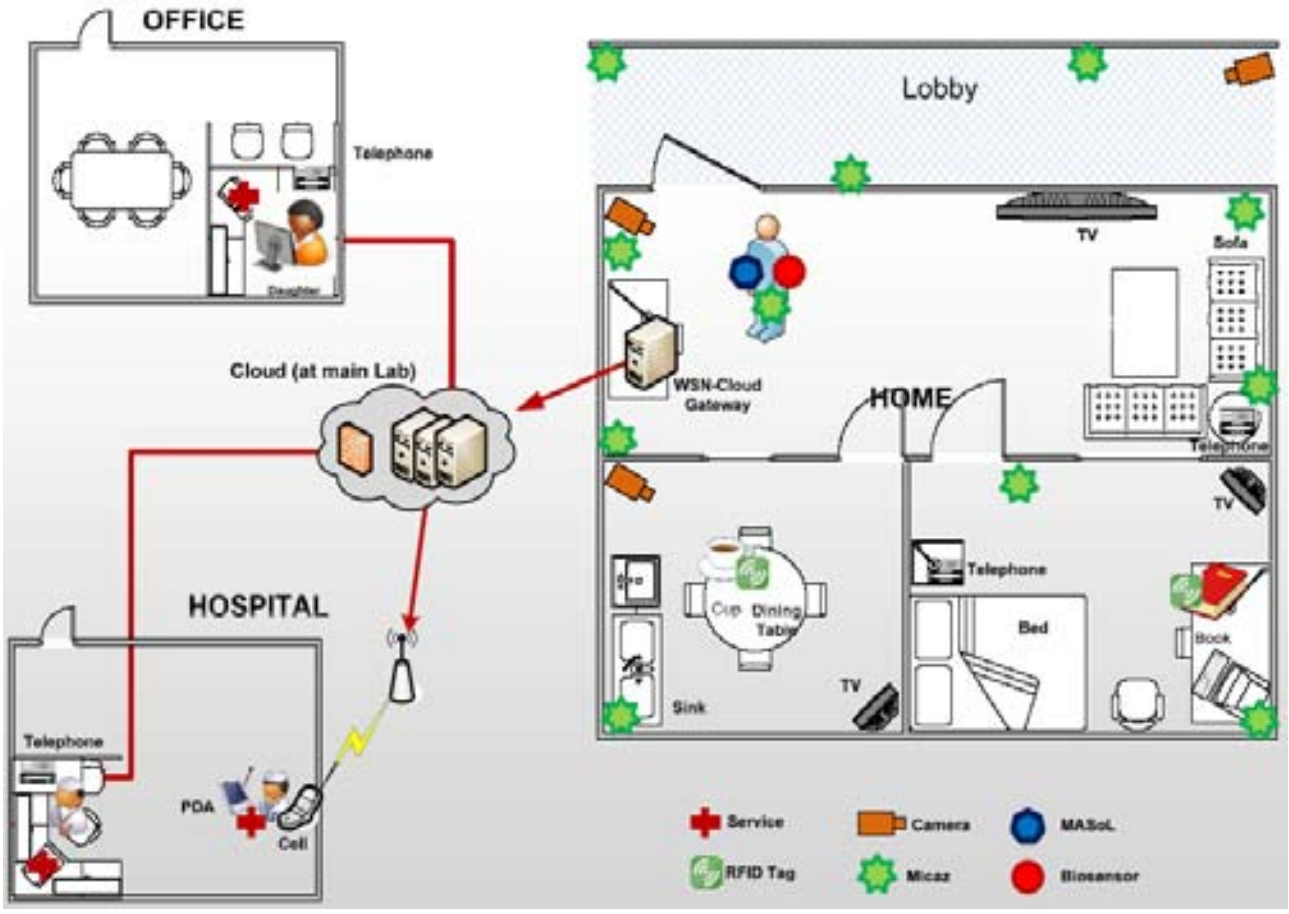

Fig. 9. A medical health care application scenario.

From Fig. 9 it can be seen that the application scenario above includes most components and models in our proposed CPeSC$^3$ architecture. For example, the wired and wireless communication is built between different systems. The computation core can be implemented both inside the cloud and the gateway (or even sensing devices). Real time scheduling and resource management can be done inside the cloud with various security strategies.

It is worth noting that different application services can be developed at hospital side to enhance the quality of health monitoring and care. In certain area at home where camera is not applicable, RFID tag or ZigBee devices like MicaZ can be utilized instead.

## 6. CONCLUSIONS AND FUTURE WORK

Cyber-physical systems are supposed to play an important role in the design of future engineering systems with more powerful capabilities than today's counterpart. Still in its infancy, CPS has many research issues and challenges. In this paper, a novel CPeSC$^3$ architecture is proposed which includes sensing, communication, computation, cloud computing and security cores. The architecture is explained in detail together with different modules such as security model, cloud computing model etc. A medical application scenario is also presented to help understanding of our proposed CPeSC$^3$ architecture.

For future research, it is the plan to focus on one or two of the components like communication and security cores inside CPeSC$^3$ architecture and try to solve some specific technical problems therein. Also, the research work can be extended by enhancing the test bed which has been built for 3 years under our 8-year governmental project.

## ACKNOWLEDGEMENT

This work was supported by a post-doctoral fellowship grant from the Kyung Hee University Korea in 2011 (KHU-20110219). This research work was also supported by the MKE (Ministry of Knowledge Economy), Korea, under the ITRC (Information Technology Research Center) support program supervised by the IITA (Institute of Information Technology Advancement) (IITA-2010-(C1090-1002-0003)) and by the basic science research program through the National Research Foundation (NRF) of Korea funded by the Ministry of Education, Science and Technology (2010-0016042). Professor Sungyoung Lee is the corresponding author. Lei Shu's research work in this paper was supported by Grant-in-Aid for Science Research (S) (21220002) of the Ministry of Education, Culture, Sports, Science and Technology, Japan. Finally, the authors are very thankful to the anonymous reviewers for their constructive comments and suggestions which greatly improve the quality of this paper.